\documentclass[sigconf,natbib=true]{acmart}

\AtBeginDocument{%
  }

\setcopyright{acmlicensed}
\copyrightyear{2026}
\acmYear{2026}
\acmDOI{XXXXXXX.XXXXXXX}

\acmISBN{978-1-4503-XXXX-X/2018/06}

\usepackage{amsmath} 
\usepackage{amsfonts} 
\usepackage{mathrsfs}
\usepackage{enumitem}
\usepackage{booktabs}  
\usepackage{tcolorbox}
\usepackage{multirow} 
\usepackage{graphicx}  
\usepackage{amsmath}  
\usepackage{mathrsfs}  
\usepackage{subcaption}
\usepackage{caption}
\usepackage{subcaption}

\usepackage{algorithm}
\usepackage{algorithmic}
\allowdisplaybreaks

\begin{document}

\title{EvoRec: Self-Evolving Agentic Recommender Systems}

\author{Lingyu Mu}
\authornotemark[1]
\affiliation{
  \institution{Alibaba International Digital Commerce Group}
  \city{Beijing} 
  \state{} 
  \country{China}
}
\email{moulingyu.mly@alibaba-inc.com}

\author{Hao Deng}
\orcid{0009-0002-6335-7405}
\authornote{Contributed equally to this research.} 
\affiliation{%
  \institution{Alibaba International Digital Commerce Group}
   \city{Beijing} 
   \state{} 
   \country{China}
}
\email{denghao.deng@alibaba-inc.com}

\author{Haibo Xing}
\orcid{0009-0006-5786-7627}
\affiliation{%
  \institution{Alibaba International Digital Commerce Group}
  \city{Hangzhou} 
  \state{} 
  \country{China}
}
\email{xinghaibo.xhb@alibaba-inc.com}

\author{Jinxin Hu}
\authornote{Corresponding authors.}
\orcid{0000-0002-7252-5207}
\affiliation{
  \institution{Alibaba International Digital Commerce Group}
  \city{Beijing} 
  \state{} 
  \country{China}
}
\email{jinxin.hjx@alibaba-inc.com}

\author{Yu Zhang}
\orcid{0000-0002-6057-7886}
\affiliation{
  \institution{Alibaba International Digital Commerce Group}
  \city{Beijing} 
  \state{} 
  \country{China}
}
\email{daoji@alibaba-inc.com}

\author{Xiaoyi Zeng}
\orcid{0000-0002-3742-4910}
\affiliation{
  \institution{Alibaba International Digital Commerce Group}
  \city{Hangzhou} 
  \state{} 
  \country{China}
}
\email{yuanhan@taobao.com}


\begin{abstract}
Optimizing modern recommender systems still relies heavily on engineers iterating by hand, which is slow and bounded by individual expertise. LLM-based agents open a path toward automating this loop, yet existing approaches use the agent only as a code translator that accumulates no methodology, and confine the search to a predefined space that rarely introduces structurally new ideas. We propose EvoRec, a multi-agent framework that co-evolves the recommendation model and the optimization methodology driving it. Four collaborating agents carry out a dual-track loop: the Research Agent and Code Agent iterate the model each round, while the Skill Evolver periodically distills reusable methodology from a persistent Memory of past experiments. Experiments on a public benchmark and one large-scale industrial dataset show that EvoRec improves offline metrics by up to 5.54\% over the strongest baseline. An online A/B test further delivers a 1.85\% revenue lift and a 1.02\% CTR gain, demonstrating the potential to replace traditional manual optimization workflows.
\end{abstract}

\vspace{-6pt}
\vspace{-6pt}


\maketitle
\section{Introduction}
Recommender systems serve as the backbone connecting users with information on Internet platforms \cite{wang2021survey,wang2024rethinking,mu2025trust}. Modern recommender systems have evolved from collaborative filtering into sophisticated multi-stage pipelines \cite{cobra,reg4rec,mu2026masked}, where each stage is shaped by extensive feature engineering, architecture design, and hyperparameter tuning \cite{morec,embedding-1,embedding-2,embedding-3,embedding-4}. This process is inefficient and bounded by individual expertise, leaving large portions of the optimization space underexplored \cite{sasrec,csmf}.

Recent advances in large language models (LLMs) have enabled agents with multi-step planning, tool invocation, and code generation capabilities to autonomously carry out complex optimization workflows much like human engineers \cite{yi2025recgpt,tang2025interactive,liu2025recoworld}. However, existing efforts still suffer from three limitations: \textbf{(1) Shallow agent involvement.} Most methods use the agent only for code generation while humans specify the optimization direction, leaving the agent's reasoning and planning capacities underutilized \cite{hao2025oxygenrec,ou2026deep}. \textbf{(2) Static experience components.} Some works introduce Skill or Memory modules \cite{packer2023memgpt,chhikara2025mem0,xu2026agent}, but these remain frozen after deployment, causing the same failures to recur and preventing successful strategies from being reused. \textbf{(3) Confined evolution space.} A few works \cite{cheng2026let,wang2026self} explore self-evolution \cite{zhang2026coevoskills,ni2026trace2skill,ma2026skillclaw}, but are restricted to hyperparameter search within a predefined space, unable to introduce novel methods.

Based on these observations, we aim to build a self-evolving recommender system that satisfies three properties: \textbf{(1) Open-domain exploration}: capable of introducing structurally new methods and insights. \textbf{(2) Human-free operation}: the entire pipeline is autonomously driven by collaborating agents. \textbf{(3) Self-improving experience}: the system continuously distills knowledge from its own evolution trajectory, growing stronger over time.

To address these challenges, we propose EvoRec, a self-evolving framework for recommender systems. EvoRec comprises three core components: a self-evolving Model, a self-evolving Skill library, and a persistent Memory store. Memory persists the complete trajectory of each iteration in structured form, including the optimization hypothesis, code modifications, training logs, evaluation metrics, and success/failure attribution. At the execution level, EvoRec employs four collaborating agents. The Orchestrator agent coordinates the overall optimization loop and drives two self-evolution pathways:

\begin{figure*}[t]

  \includegraphics[width=1.0\textwidth]{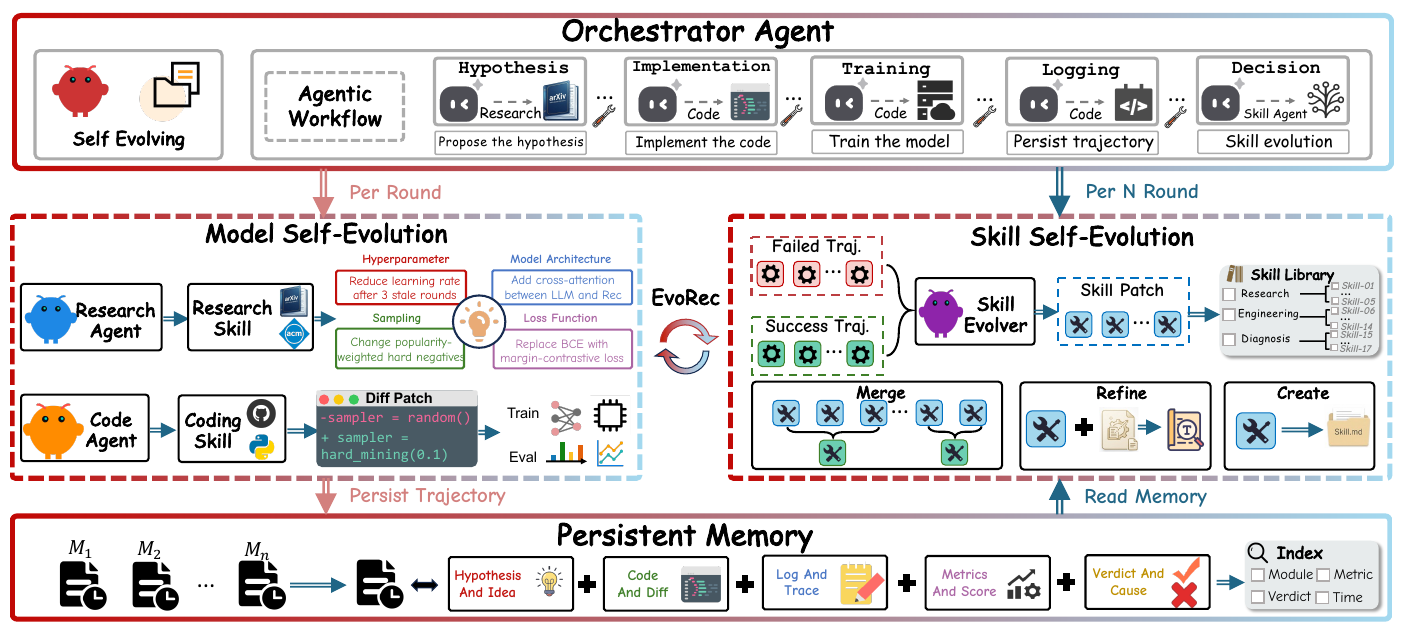}
  \vspace{-10pt}
    \caption{The overview of EvoRec. Four collaborating agents drive dual-track self-evolution: the Research Agent and Code Agent perform per-round model self-evolution, while the Skill Evolver periodically distills methodologies from persistent Memory to update the Skill library.}

  \label{figure1}
\end{figure*}

\begin{itemize}[noitemsep, topsep=0pt, leftmargin=*]
    \item (1) Model self-evolution is driven by the Research Agent and Code Agent. The Research Agent retrieves external knowledge (e.g., arXiv) and combines it with Skill and Memory to generate optimization hypotheses. The Code Agent translates each hypothesis into code edits and submits training.
    \item (2) Skill self-evolution is handled by the Skill Evolver, which distills reusable methodologies from Memory every $N$ rounds and updates the Skill library.
\end{itemize}
The model iterates through a hypothesis–implementation–validation cycle each round, while Skills evolve periodically after sufficient experience has accumulated. Under the coordination of the Orchestrator, the two pathways are coupled through persistent Memory, forming a self-reinforcing loop in which the model grows stronger and the methodologies become increasingly refined. Experiments on a public benchmark and one industrial dataset show up to 5.54\% offline improvement over the strongest baselines, and an online A/B test confirms a 1.85\% revenue lift and 1.02\% CTR gain.

Our contributions can be summarized as follows:
\begin{itemize}[noitemsep, topsep=0pt, leftmargin=*]
    \item We propose EvoRec, the first multi-agent self-evolving optimization framework for recommender systems, enabling continuous model self-evolution without human intervention.
    \item We design a dual-track self-evolution mechanism driven by four collaborating agents, achieving co-evolution of Model and Skill through persistent Memory.
    \item We validate the effectiveness of EvoRec on a public benchmark and an industrial advertising platform, and further demonstrate its practical deployment value through online A/B testing.
\end{itemize}

\section{Method}
\subsection{Overview}
As illustrated in Figure~\ref{figure1}, EvoRec consists of three core components: the Model, an evolvable Skill library, and a structured Memory, coordinated by four collaborating agents. Each iteration sequentially completes hypothesis generation, code implementation, training, and experience persistence, forming a dual-track loop where the model and methodologies evolve in tandem.

\subsection{Orchestrator Agent}
The Orchestrator maintains global state and coordinates the other three agents. At each iteration, it activates the Research Agent and Code Agent to complete one round of model optimization, then writes the full trajectory into Memory. Every $N$ rounds, it triggers the Skill Evolver to update the Skill library. The loop terminates when reaching $T_{\max}$ iterations, $k$ consecutive rounds without improvement, or a target metric threshold. Failed rounds are preserved in Memory as negative examples rather than discarded.

\subsection{Model Self-Evolution}

Model self-evolution is jointly carried out by the Research Agent and the Code Agent, forming the core of each EvoRec iteration. Let the global state at round $t$ be $\langle \theta_{t-1}, \mathcal{S}_t, \mathcal{M}_t \rangle$, denoting the current model parameters, the Skill library, and the Memory experience store, respectively.

\subsubsection{Research Agent.}
The Research Agent is responsible for generating the optimization hypothesis for the current round. It retrieves candidate research directions from external knowledge sources $\mathcal{K}_{\text{ext}}$ (e.g., arXiv, technical blogs), weighs them against the Research-type methodologies in $\mathcal{S}_t$ and the historical success and failure cases in $\mathcal{M}_t$, and outputs:
\begin{equation}
    h_t \sim \pi_R(\cdot \mid \mathcal{K}_{\text{ext}}, \mathcal{S}_t, \mathcal{M}_t),
\end{equation}
where $h_t = \langle \textit{motive}, \textit{target\_module}, \textit{expected\_gain} \rangle$ comprises the optimization motive, the target module to modify, and the expected gain. Before generating a hypothesis, the Research Agent performs a diversity constraint check to avoid focusing on the same module for too many consecutive rounds, which could lead to local optima. Retrieval over $\mathcal{M}_t$ allows it to steer clear of previously failed directions, while injection of $\mathcal{S}_t$ biases it toward exploration patterns that have proven effective.

\subsubsection{Code Agent.}
The Code Agent translates the hypothesis $h_t$ into an executable experiment. It locates the relevant code, invokes Engineering-type methodologies in $\mathcal{S}_t$ to generate minimal-diff modifications, and submits the training job. Upon completion, it outputs the round result $r_t$ containing training logs and evaluation metrics. If $h_t$ is deemed infeasible during implementation, a rejection signal is returned and the Research Agent generates an alternative hypothesis. Failed runs are marked with failure reasons and preserved for subsequent Skill evolution.

\subsection{Memory: Persistent Experience Storage}

Memory is the experience hub of EvoRec. It persists the complete trajectory of each iteration in structured form, serving both as a retrieval source for the Research Agent and as distillation material for the Skill Evolver. After each iteration, the Orchestrator packages the round's output into a memory entry:
\begin{equation}
    m_t = \langle h_t, \Delta c_t, \textit{log}_t, \textit{metric}_t, \textit{verdict}_t \rangle,
\end{equation}
where $h_t$ is the optimization hypothesis, $\Delta c_t$ the code modifications, $\textit{log}_t$ the training log summary, $\textit{metric}_t$ the offline evaluation metric vector, and $\textit{verdict}_t \in \{\text{success}, \text{fail}, \text{neutral}\}$ the attribution label. All entries are appended chronologically to form $\mathcal{M}_t = \{m_1, m_2, \dots, m_t\}$.

To support efficient access by upstream agents, Memory builds a structured index upon each write. Specifically, metadata fields are extracted from each $m_t$, including \textit{target\_module} (the modified model module), \textit{verdict} (the attribution label), \textit{metric\_delta} (the metric change relative to the previous version), and \textit{timestamp}. These fields are stored as an inverted index that supports exact field-level filtering. Memory itself does not participate in self-evolution and serves purely as passive storage.

\subsection{Skill Self-Evolution}

Unlike the per-round experiment trajectories stored in Memory, Skills are robust operational principles produced by cross-round inductive compression rather than local experience. They are expressed in declarative natural language, not tied to any specific model architecture, and can be reused across different optimization objectives. Formally, a Skill is defined as a four-tuple $s = \langle \textit{name}, \textit{scope}, \textit{content}, \textit{evidence} \rangle$, where \textit{name} is the skill identifier, \textit{scope} specifies the applicable conditions (e.g., ``attention selection'' or ``loss not decreasing''), \textit{content} provides the concrete operational guidelines whose granularity ranges from a single rule of thumb to a multi-step conditional procedure and grows richer as evolution progresses, and \textit{evidence} points back to the Memory entries that support the rule to ensure traceability. A short example is:
\begin{tcolorbox}[colback=gray!5, colframe=gray!60, boxrule=0.5pt, left=4pt, right=4pt, top=3pt, bottom=3pt, fontupper=\small]
\textbf{Skill \#12}: Hard Negative Mining\\[2pt]
\textit{Scope}: Sampling strategy when recall plateaus\\
\textit{Content}: When R@10 stagnates for $\geq$3 rounds, replace random negatives with popularity-weighted hard negatives ($\tau \in [0.05, 0.15]$). Start with $\tau{=}0.1$ and adjust based on training loss variance.\\
\textit{Evidence}: Memory entries $m_{14}, m_{18}, m_{22}$
\label{skill}
\end{tcolorbox}

\noindent By functional stage, Skills are divided into three categories: Research Skills (guiding hypothesis formulation), Engineering Skills (guiding code implementation), and Diagnosis Skills (guiding failure repair, distilled primarily from $\textit{verdict}{=}\text{fail}$ cases). The Orchestrator matches each Skill's \textit{scope} against the current objective and injects only relevant entries into the agent's context. All categories share the same evolution pipeline, collectively forming $\mathcal{S}_t$.

\subsubsection{Triggering and Batch Selection.}
Skill self-evolution is carried out by the Skill Evolver agent and triggered by the Orchestrator every $N$ rounds of Model iteration. At the triggering round $t$, the Skill Evolver selects the most recent $N$ rounds of experiment records from Memory as the distillation batch:
\begin{equation}
    \mathcal{B}_t = \{ m_\tau \in \mathcal{M}_t \mid t - N + 1 \leq \tau \leq t \},
\end{equation}
and applies balanced filtering by \textit{verdict} label so that success, fail, and neutral cases appear in the batch at a preset ratio $\rho$. This balance ensures that the distillation process extracts positive patterns from successful cases while also consolidating lessons from failures.

\subsubsection{Distillation and Update.}
Given batch $\mathcal{B}_t$ and the current Skill library $\mathcal{S}_t$, the Skill Evolver generates a set of candidate Skill patches via policy $\pi_E$:
\begin{equation}
    \mathcal{C}_t = \{ c_1, \dots, c_K \} \sim \pi_E(\cdot \mid \mathcal{B}_t, \mathcal{S}_t).
\end{equation}
Each candidate $c_k$ follows the same four-tuple structure. For each candidate, the system computes its maximum similarity to the existing Skill library $\sigma_k = \max_{s \in \mathcal{S}_t} \text{sim}(c_k, s)$, and applies one of three update operators based on threshold $\tau_c$:

\begin{itemize}[noitemsep, topsep=0pt, leftmargin=*]
    \item \textbf{Merge}: Candidates that are pairwise similar above $\tau_c$ within the batch are first consolidated into a single candidate, preventing redundant Skills from being introduced simultaneously.
    \item \textbf{Refine}: When $\sigma_k \geq \tau_c$, $c_k$ is merged with its most similar existing Skill and revised, corresponding to methodology refinement.
    \item \textbf{Create}: When $\sigma_k < \tau_c$, candidate $c_k$ is added to the library as a new Skill, corresponding to methodology expansion.
\end{itemize}

The Skill library is then updated as:
\begin{equation}
    \mathcal{S}_{t+1} = (\mathcal{S}_t \setminus \mathcal{S}_{\text{refined}}) \cup \mathcal{S}_{\text{refined}}' \cup \mathcal{S}_{\text{new}}.
\end{equation}

\begin{table*}[t]
\caption{Performance comparison on the Books and Industrial datasets. ``Imp.'' shows the relative improvement (\%) over the strongest baseline in each group. Best results per group are in \textbf{bold} and second-best are \underline{underlined}.}
\label{tab:main}
\centering
\resizebox{\textwidth}{!}{
\begin{tabular}{ll ccccc cc ccccc cc}
\toprule
\multirow{2}{*}{} & \multirow{2}{*}{} & \multicolumn{7}{c}{\textbf{\textit{ID-based Recommendation}}} & \multicolumn{7}{c}{\textbf{\textit{Generative Recommendation}}} \\
\cmidrule(lr){3-9} \cmidrule(lr){10-16}
 & & SASRec & S$^3$Rec & PFormer & MGUI & Auto-MGUI & Evo-MGUI & Imp. & HSTU & TIGER & Cobra & REG4Rec & Auto-REG & Evo-REG4Rec & Imp. \\
\midrule
\multirow{4}{*}{Books}
 & R@5  & 0.0472 & 0.0471 & 0.0492 & 0.0503 & \underline{0.0506} & \textbf{0.0527} & \textbf{4.15\%} & 0.0536 & 0.0562 & 0.0584 & 0.0587 & \underline{0.0592} & \textbf{0.0608} & \textbf{2.70\%} \\
 & N@5  & 0.0267 & 0.0264 & 0.0281 & \underline{0.0284} & 0.0279 & \textbf{0.0298} & \textbf{4.93\%} & 0.0307 & 0.0324 & \underline{0.0335} & 0.0331 & 0.0327 & \textbf{0.0349} & \textbf{4.18\%} \\
 & R@10 & 0.0599 & 0.0597 & 0.0623 & \underline{0.0635} & 0.0630 & \textbf{0.0664} & \textbf{4.57\%} & 0.0697 & 0.0733 & 0.0761 & \underline{0.0768} & 0.0763 & \textbf{0.0792} & \textbf{3.13\%} \\
 & N@10 & 0.0343 & 0.0339 & 0.0357 & \underline{0.0361} & 0.0358 & \textbf{0.0381} & \textbf{5.54\%} & 0.0392 & 0.0415 & 0.0429 & 0.0425 & \underline{0.0431} & \textbf{0.0446} & \textbf{3.48\%} \\
\midrule
\multirow{4}{*}{Industry}
 & R@5  & 0.0701 & 0.0706 & 0.0734 & 0.0741 & \underline{0.0745} & \textbf{0.0775} & \textbf{4.03\%} & 0.0832 & 0.0904 & 0.0965 & \underline{0.1016} & 0.1008 & \textbf{0.1053} & \textbf{3.64\%} \\
 & N@5  & 0.0919 & 0.0939 & \underline{0.0984} & 0.0982 & 0.0976 & \textbf{0.1037} & \textbf{5.39\%} & 0.1062 & 0.1165 & 0.1236 & 0.1294 & \underline{0.1302} & \textbf{0.1331} & \textbf{2.23\%} \\
 & R@10 & 0.0518 & 0.0523 & 0.0544 & \underline{0.0547} & 0.0542 & \textbf{0.0571} & \textbf{4.39\%} & 0.0613 & 0.0672 & 0.0701 & \underline{0.0736} & 0.0731 & \textbf{0.0761} & \textbf{3.40\%} \\
 & N@10 & 0.0681 & 0.0697 & 0.0741 & 0.0736 & \underline{0.0743} & \textbf{0.0776} & \textbf{4.44\%} & 0.0811 & 0.0871 & 0.0916 & 0.0957 & \underline{0.0961} & \textbf{0.0993} & \textbf{3.33\%} \\
\bottomrule
\end{tabular}
}
\end{table*}

\section{Experiment}
\subsection{Experimental Setup}

\subsubsection{Datasets and Metrics.}
We evaluate on one public dataset (Amazon Books \cite{AmazonDataset}, filtered to users with $\geq$5 interactions) and one industrial dataset from the internal interaction logs of a Southeast Asian e-commerce platform spanning January to May 2026, containing approximately 2.95B user-item interactions, 14.7M users, and 20.1M items. We adopt leave-one-out evaluation and report Recall@5/10 (R@5/10) and NDCG@5/10 (N@5/10) \cite{reg4rec}.

\subsubsection{Baselines and Base Models.}
We compare against \textit{ID-based recommenders} (SASRec \cite{sasrec}, S$^3$Rec \cite{zhou2020s3}, PinnerFormer \cite{pancha2022pinnerformer}), \textit{generative recommenders} (HSTU \cite{zhai2024actions}, TIGER \cite{tiger}, Cobra \cite{cobra}), and \textit{auto-optimization} (AutoML \cite{he2021automl}, applied to each base model respectively as Auto-MGUI and Auto-REG). We select MGUI \cite{wu2026modeling} (ID-based) and REG4Rec \cite{xing2025reg4rec} (generative) as base models for EvoRec, yielding Evo-MGUI and Evo-REG4Rec.

\subsubsection{Implementation Details.}
All models are trained with PyTorch on 4 NVIDIA A100 GPUs \cite{paszke2019pytorch}. The four agents share Claude Opus 4.6 as the reasoning backbone, with $N{=}5$, $T_{\max}{=}50$, $k{=}10$, $\tau_c{=}0.6$, and $\rho{=}(0.5, 0.3, 0.2)$. Training hyperparameters for both base models follow their original papers. Reproducing the pipeline with Qwen3.7-MAX \cite{bai2023qwen} yields consistent conclusions, confirming robustness to the LLM backbone choice.

\subsection{Main Results}

Table~\ref{tab:main} reports results on the public and industrial datasets. EvoRec-MGUI and EvoRec-REG4Rec achieve the best performance across all datasets and metrics, with up to 5.54\% relative improvement over the strongest baseline. The consistent gains across both ID-based and generative paradigms confirm the generality of the framework. The improvement on the industrial dataset is generally larger than on Books, as the richer feature space provides a wider optimization landscape. Meanwhile, gains in the generative paradigm are slightly smaller, as generative models already incorporate stronger representational priors. Compared with AutoML, which is bounded by a fixed configuration space, EvoRec achieves consistent advantages because its Skill self-evolution mechanism accumulates and refines optimization methodologies across iterations, continuously identifying better directions beyond predefined search ranges.

\subsection{Ablation Study}

To verify the contribution of each core component, we conduct three ablation experiments on the industrial dataset using EvoRec-MGUI as the base, with results reported in Table~\ref{tab:ablation}.

\begin{table}[t]
\caption{Ablation study on the Industrial dataset using EvoRec-MGUI. $\Delta$ denotes the relative drop in R@10 compared to the full model.}
\label{tab:ablation}
\centering
\small
\begin{tabular}{l cccc c}
\toprule
Method & R@5 & N@5 & R@10 & N@10 & $\Delta$R@10 \\
\midrule
EvoRec-MGUI & \textbf{0.0775} & \textbf{0.1037} & \textbf{0.0571} & \textbf{0.0776} & -- \\
\midrule
w/o Skill Evo. & 0.0759 & 0.1015 & 0.0559 & 0.0761 & -2.10\% \\
w/o Memory & 0.0749 & 0.1003 & 0.0551 & 0.0750 & -3.50\% \\
w/o Ext. Know. & 0.0763 & 0.1021 & 0.0562 & 0.0764 & -1.58\% \\
\bottomrule
\end{tabular}
\end{table}

\begin{itemize}[noitemsep, topsep=0pt, leftmargin=*]
    \item \textbf{w/o Skill Evolution} ($-$2.10\% R@10): The Skill library stays frozen. The drop confirms that continuously refined methodologies are critical for guiding later-stage optimization.
    \item \textbf{w/o Memory} ($-$3.50\% R@10): Without historical records, the system repeatedly explores failed directions, confirming Memory's role in accelerating convergence.
    \item \textbf{w/o External Knowledge} ($-$1.58\% R@10): Relying solely on internal experience limits the system to known patterns, showing that external knowledge introduces structurally novel solutions.
\end{itemize}

\subsection{Case Study}

We use the optimization trajectory of EvoRec-MGUI on the industrial dataset to illustrate the typical working behavior of EvoRec across 50 iterations. Table~\ref{tab:case} lists the rounds with the most significant metric changes, including both successful and failed attempts. 

\begin{table}[t]
\caption{Rounds with the most significant R@10 changes during EvoRec-MGUI optimization on the Industrial dataset. }
\label{tab:case}
\centering
\resizebox{\columnwidth}{!}{
\begin{tabular}{c l l c l}
\toprule
\textbf{Rnd} & \textbf{Category} & \textbf{Modification} & \textbf{$\Delta$R@10} & \textbf{Source} \\
\midrule
3  & Hyperparam     & Reduce lr to 5e-4                & +0.41\% & Research \\
11 & Loss           & Focal loss variant               & -0.82\% & Research \\
14 & Sampling       & Neg. samples 4$\to$16            & +0.52\% & Research \\
19 & Architecture   & Multi-interest pooling (4 heads)  & +1.03\% & Research \\
24 & Loss           & Replace BCE with InfoNCE          & +1.18\% & Research \\
27 & Sampling       & Hard neg. mining ($\tau{=}0.1$)   & \textbf{+2.31\%} & Skill \#12 \\
33 & Hyperparam     & Layer-wise lr decay               & +0.63\% & Skill \#16 \\
36 & Architecture   & Attention head pruning            & -0.87\% & Research \\
\bottomrule
\end{tabular}
}
\end{table}

The trajectory spans diverse categories including hyperparameters, sampling, architecture, and loss design. Not all attempts succeed: rounds 11 and 36 cause regressions but are preserved in Memory as negative examples, helping the Research Agent avoid similar directions later. After sufficient experience accumulates, Skill-driven rounds emerge (27, 33) and both succeed, with round 27 achieving the largest single-round gain (+2.31\%) via Skill \#12 distilled from prior sampling experiments. This suggests that distilled methodologies provide more reliable directions than open-ended exploration, while the two sources play complementary roles.
\subsection{Online A/B Test}

We deploy EvoRec-MGUI on the advertising recommendation platform of a leading e-commerce company in Southeast Asia for a 7-day online A/B test. The control group uses the production model MGUI, and the experimental group replaces it with the EvoRec-optimized
variant. Each group contains 20\% of users sampled uniformly at random. EvoRec-MGUI delivers a \textbf{1.85\%} lift in advertising revenue and a \textbf{1.02\%} improvement in CTR, both statistically significant under a two-sided test ($p < 0.05$). These results confirm that the offline gains of EvoRec translate into real business value.

\section{Conclusion}

In this work, we propose EvoRec, a multi-agent self-evolving framework for recommender systems that replaces manual optimization with autonomous agent collaboration. We identify three limitations of prior agent-based approaches: shallow involvement limited to code translation, static experience that never updates, and evolution confined to predefined search spaces. EvoRec addresses these via a dual-track mechanism where the Model and the Skill library co-evolve through persistent Memory. Extensive offline and online experiments validate the effectiveness of EvoRec across both public and industrial settings. Future work includes multi-stage cascade optimization and cross-task Skill transfer across different recommendation scenarios.


\bibliographystyle{ACM-Reference-Format}
\bibliography{main}

@inproceedings{AmazonDataset,
author = {He, Ruining and McAuley, Julian},
title = {Ups and Downs: Modeling the Visual Evolution of Fashion Trends with One-Class Collaborative Filtering},
year = {2016},
isbn = {9781450341431},
publisher = {International World Wide Web Conferences Steering Committee},
address = {Republic and Canton of Geneva, CHE},
url = {https://doi.org/10.1145/2872427.2883037},
doi = {10.1145/2872427.2883037},
booktitle = {Proceedings of the 25th International Conference on World Wide Web},
pages = {507–517},
numpages = {11},
location = {Montr\'{e}al, Qu\'{e}bec, Canada},
series = {WWW '16}
}

@article{wang2021survey,
  title={A survey on session-based recommender systems},
  author={Wang, Shoujin and Cao, Longbing and Wang, Yan and Sheng, Quan Z and Orgun, Mehmet A and Lian, Defu},
  journal={ACM Computing Surveys (CSUR)},
  volume={54},
  number={7},
  pages={1--38},
  year={2021},
  publisher={ACM New York, NY, USA}
}

@article{cheng2026let,
  title={Let the Agent Steer: Closed-Loop Ranking Optimization via Influence Exchange},
  author={Cheng, Yin and Zhou, Liao and Liang, Xiyu and Luo, Dihao and Lee, Tewei and Zheng, Kailun and Zhang, Weiwei and Cai, Mingchen and Dong, Jian and Zhang, Andy},
  journal={arXiv preprint arXiv:2603.27765},
  year={2026}
}

@article{xu2026agent,
  title={Agent skills for large language models: Architecture, acquisition, security, and the path forward},
  author={Xu, Renjun and Yan, Yang},
  journal={arXiv preprint arXiv:2602.12430},
  year={2026}
}

@article{packer2023memgpt,
  title={MemGPT: towards LLMs as operating systems.},
  author={Packer, Charles and Fang, Vivian and Patil, Shishir\_G and Lin, Kevin and Wooders, Sarah and Gonzalez, Joseph\_E},
  year={2023},
  publisher={ArXiv}
}

@article{zhang2026coevoskills,
  title={Coevoskills: Self-evolving agent skills via co-evolutionary verification},
  author={Zhang, Hanrong and Fan, Shicheng and Zou, Henry Peng and Chen, Yankai and Wang, Zhenting and Zhou, Jiayu and Li, Chengze and Huang, Wei-Chieh and Yao, Yifei and Zheng, Kening and others},
  journal={arXiv preprint arXiv:2604.01687},
  year={2026}
}

@article{he2021automl,
  title={AutoML: A survey of the state-of-the-art},
  author={He, Xin and Zhao, Kaiyong and Chu, Xiaowen},
  journal={Knowledge-based systems},
  volume={212},
  pages={106622},
  year={2021},
  publisher={Elsevier}
}

@inproceedings{zhou2020s3,
  title={S3-rec: Self-supervised learning for sequential recommendation with mutual information maximization},
  author={Zhou, Kun and Wang, Hui and Zhao, Wayne Xin and Zhu, Yutao and Wang, Sirui and Zhang, Fuzheng and Wang, Zhongyuan and Wen, Ji-Rong},
  booktitle={Proceedings of the 29th ACM international conference on information \& knowledge management},
  pages={1893--1902},
  year={2020}
}

@article{paszke2019pytorch,
  title={Pytorch: An imperative style, high-performance deep learning library},
  author={Paszke, Adam and Gross, Sam and Massa, Francisco and Lerer, Adam and Bradbury, James and Chanan, Gregory and Killeen, Trevor and Lin, Zeming and Gimelshein, Natalia and Antiga, Luca and others},
  journal={Advances in neural information processing systems},
  volume={32},
  year={2019}
}

@article{xing2025reg4rec,
  title={Reg4rec: Reasoning-enhanced generative model for large-scale recommendation systems},
  author={Xing, Haibo and Deng, Hao and Mao, Yucheng and Hu, Jinxin and Xu, Yi and Zhang, Hao and Wang, Jiahao and Wang, Shizhun and Zhang, Yu and Zeng, Xiaoyi and others},
  journal={arXiv preprint arXiv:2508.15308},
  year={2025}
}

@article{ni2026trace2skill,
  title={Trace2skill: Distill trajectory-local lessons into transferable agent skills},
  author={Ni, Jingwei and Liu, Yihao and Liu, Xinpeng and Sun, Yutao and Zhou, Mengyu and Cheng, Pengyu and Wang, Dexin and Zhao, Erchao and Jiang, Xiaoxi and Jiang, Guanjun},
  journal={arXiv preprint arXiv:2603.25158},
  year={2026}
}

@article{ma2026skillclaw,
  title={Skillclaw: Let skills evolve collectively with agentic evolver},
  author={Ma, Ziyu and Yang, Shidong and Ji, Yuxiang and Wang, Xucong and Wang, Yong and Hu, Yiming and Huang, Tongwen and Chu, Xiangxiang},
  journal={arXiv preprint arXiv:2604.08377},
  year={2026}
}

@article{wang2026self,
  title={Self-evolving recommendation system: End-to-end autonomous model optimization with LLM agents},
  author={Wang, Haochen and Wu, Yi and Chang, Daryl and Wei, Li and Heldt, Lukasz},
  journal={arXiv preprint arXiv:2602.10226},
  year={2026}
}

@article{chhikara2025mem0,
  title={Mem0: Building production-ready ai agents with scalable long-term memory},
  author={Chhikara, Prateek and Khant, Dev and Aryan, Saket and Singh, Taranjeet and Yadav, Deshraj},
  journal={arXiv preprint arXiv:2504.19413},
  year={2025}
}

@article{wang2024rethinking,
  title={Rethinking large language model architectures for sequential recommendations},
  author={Wang, Hanbing and Liu, Xiaorui and Fan, Wenqi and Zhao, Xiangyu and Kini, Venkataramana and Yadav, Devendra and Wang, Fei and Wen, Zhen and Tang, Jiliang and Liu, Hui},
  journal={arXiv preprint arXiv:2402.09543},
  year={2024}
}

@article{bai2023qwen,
  title={Qwen technical report},
  author={Bai, Jinze and Bai, Shuai and Chu, Yunfei and Cui, Zeyu and Dang, Kai and Deng, Xiaodong and Fan, Yang and Ge, Wenbin and Han, Yu and Huang, Fei and others},
  journal={arXiv preprint arXiv:2309.16609},
  year={2023}
}

@inproceedings{morec,
  title={Where to go next for recommender systems? id-vs. modality-based recommender models revisited},
  author={Yuan, Zheng and Yuan, Fajie and Song, Yu and Li, Youhua and Fu, Junchen and Yang, Fei and Pan, Yunzhu and Ni, Yongxin},
  booktitle={Proceedings of the 46th International ACM SIGIR Conference on Research and Development in Information Retrieval},
  pages={2639--2649},
  year={2023}
}

@inproceedings{embedding-1,
  title={Empowering news recommendation with pre-trained language models},
  author={Wu, Chuhan and Wu, Fangzhao and Qi, Tao and Huang, Yongfeng},
  booktitle={Proceedings of the 44th international ACM SIGIR conference on research and development in information retrieval},
  pages={1652--1656},
  year={2021}
}

@inproceedings{embedding-2,
  title={Boosting deep CTR prediction with a plug-and-play pre-trainer for news recommendation},
  author={Liu, Qijiong and Zhu, Jieming and Dai, Quanyu and Wu, Xiao-Ming},
  booktitle={Proceedings of the 29th International Conference on Computational Linguistics},
  pages={2823--2833},
  year={2022}
}

@inproceedings{embedding-3,
  title={CTR-BERT: Cost-effective knowledge distillation for billion-parameter teacher models},
  author={Muhamed, Aashiq and Keivanloo, Iman and Perera, Sujan and Mracek, James and Xu, Yi and Cui, Qingjun and Rajagopalan, Santosh and Zeng, Belinda and Chilimbi, Trishul},
  booktitle={NeurIPS Efficient Natural Language and Speech Processing Workshop},
  year={2021}
}

@inproceedings{embedding-4,
  title={GBERT: Pre-training user representations for ephemeral group recommendation},
  author={Zhang, Song and Zheng, Nan and Wang, Danli},
  booktitle={Proceedings of the 31st ACM international conference on information \& knowledge management},
  pages={2631--2639},
  year={2022}
}

@article{wu2026modeling,
  title={Modeling Multi-Grained User Interests for Sequential Recommendation},
  author={Wu, Bin and Yin, Xiaowen and Su, Xun and Xu, Mingliang},
  journal={IEEE Transactions on Computational Social Systems},
  year={2026},
  publisher={IEEE}
}

@article{mu2026masked,
  title={Masked Diffusion Generative Recommendation},
  author={Mu, Lingyu and Deng, Hao and Xing, Haibo and Hu, Jinxin and Zhang, Yu and Zeng, Xiaoyi and Zhang, Jing},
  journal={arXiv preprint arXiv:2601.19501},
  year={2026Reg4rec: Reasoning-enhanced generative model for large-scale recommendation systems}
}

@article{tiger,
  title={Recommender systems with generative retrieval},
  author={Rajput, Shashank and Mehta, Nikhil and Singh, Anima and Hulikal Keshavan, Raghunandan and Vu, Trung and Heldt, Lukasz and Hong, Lichan and Tay, Yi and Tran, Vinh and Samost, Jonah and others},
  journal={Advances in Neural Information Processing Systems},
  volume={36},
  pages={10299--10315},
  year={2023}
}

@article{cobra,
  title={Sparse meets dense: Unified generative recommendations with cascaded sparse-dense representations},
  author={Yang, Yuhao and Ji, Zhi and Li, Zhaopeng and Li, Yi and Mo, Zhonglin and Ding, Yue and Chen, Kai and Zhang, Zijian and Li, Jie and Li, Shuanglong and others},
  journal={arXiv preprint arXiv:2503.02453},
  year={2025}
}

@article{zhai2024actions,
  title={Actions speak louder than words: Trillion-parameter sequential transducers for generative recommendations},
  author={Zhai, Jiaqi and Liao, Lucy and Liu, Xing and Wang, Yueming and Li, Rui and Cao, Xuan and Gao, Leon and Gong, Zhaojie and Gu, Fangda and He, Michael and others},
  journal={arXiv preprint arXiv:2402.17152},
  year={2024}
}

@inproceedings{pancha2022pinnerformer,
  title={Pinnerformer: Sequence modeling for user representation at pinterest},
  author={Pancha, Nikil and Zhai, Andrew and Leskovec, Jure and Rosenberg, Charles},
  booktitle={Proceedings of the 28th ACM SIGKDD conference on knowledge discovery and data mining},
  pages={3702--3712},
  year={2022}
}

@inproceedings{csmf,
  title={CSMF: Cascaded Selective Mask Fine-Tuning for Multi-Objective Embedding-Based Retrieval},
  author={Deng, Hao and Xing, Haibo and Matsuyama, Kanefumi and Zhang, Moyu and Hu, Jinxin and Wen, Hong and Zhang, Yu and Zeng, Xiaoyi and Zhang, Jing},
  booktitle={Proceedings of the 48th International ACM SIGIR Conference on Research and Development in Information Retrieval},
  pages={2122--2131},
  year={2025}
}

@article{tang2025interactive,
  title={Interactive Recommendation Agent with Active User Commands},
  author={Tang, Jiakai and Luo, Yujie and Xi, Xunke and Sun, Fei and Feng, Xueyang and Dai, Sunhao and Yi, Chao and Chen, Dian and Gao, Zhujin and Li, Yang and others},
  journal={arXiv preprint arXiv:2509.21317},
  year={2025}
}

@article{yi2025recgpt,
  title={RecGPT-V2 Technical Report},
  author={Yi, Chao and Chen, Dian and Guo, Gaoyang and Tang, Jiakai and Wu, Jian and Yu, Jing and Zhang, Mao and Chen, Wen and Yang, Wenjun and Luo, Yujie and others},
  journal={arXiv preprint arXiv:2512.14503},
  year={2025}
}

@article{ou2026deep,
  title={Deep Research for Recommender Systems},
  author={Ou, Kesha and Wu, Chenghao and Wang, Xiaolei and Zheng, Bowen and Zhao, Wayne Xin and Li, Weitao and Zhang, Long and Chen, Sheng and Wen, Ji-Rong},
  journal={arXiv preprint arXiv:2603.07605},
  year={2026}
}

@article{hao2025oxygenrec,
  title={OxygenREC: An Instruction-Following Generative Framework for E-commerce Recommendation},
  author={Hao, Xuegang and Zhang, Ming and Li, Alex and Qian, Xiangyu and Ma, Zhi and Zang, Yanlong and Yang, Shijie and Han, Zhongxuan and Ma, Xiaolong and Liu, Jinguang and others},
  journal={arXiv preprint arXiv:2512.22386},
  year={2025}
}

@article{liu2025recoworld,
  title={Recoworld: Building simulated environments for agentic recommender systems},
  author={Liu, Fei and Lin, Xinyu and Yu, Hanchao and Wu, Mingyuan and Wang, Jianyu and Zhang, Qiang and Zhao, Zhuokai and Xia, Yinglong and Zhang, Yao and Li, Weiwei and others},
  journal={arXiv preprint arXiv:2509.10397},
  year={2025}
}

@inproceedings{mu2025trust,
  title={Trust-GRS: A Trustworthy Training Framework for Graph Neural Network Based Recommender Systems Against Shilling Attacks},
  author={Mu, Lingyu and Liu, Zhengxiao and Zhu, Zhitong and Lin, Zheng},
  booktitle={Proceedings of the AAAI Conference on Artificial Intelligence},
  volume={39},
  number={12},
  pages={12408--12416},
  year={2025}
}

@inproceedings{sasrec,
  title={Self-attentive sequential recommendation},
  author={Kang, Wang-Cheng and McAuley, Julian},
  booktitle={2018 IEEE international conference on data mining (ICDM)},
  pages={197--206},
  year={2018},
  organization={IEEE}
}

@article{reg4rec,
  title={Reg4rec: Reasoning-enhanced generative model for large-scale recommendation systems},
  author={Xing, Haibo and Deng, Hao and Mao, Yucheng and Hu, Jinxin and Xu, Yi and Zhang, Hao and Wang, Jiahao and Wang, Shizhun and Zhang, Yu and Zeng, Xiaoyi and others},
  journal={arXiv preprint arXiv:2508.15308},
  year={2025}
}
\end{document}